\documentclass[12pt]{article}
\pdfoutput=1
\usepackage{graphicx,amsfonts,amsthm,amsmath,amssymb,hyperref,color,yfonts,cite}
\newcommand{\bea}{\begin{eqnarray}\displaystyle}
\newcommand{\eea}{\end{eqnarray}}

\begin{document}
\makeatletter
\@addtoreset{equation}{section}
\makeatother
\renewcommand{\theequation}{\thesection.\arabic{equation}}
\vspace{1.8truecm}

{\LARGE{ \centerline{\bf Bilocal holography and locality in the bulk}}}  

\vskip.5cm 

\thispagestyle{empty} 
\centerline{{\large\bf Robert de Mello Koch$^{a,b,e}$\footnote{{\tt robert@zjhu.edu.cn}}, Garreth Kemp$^{c,e}$\footnote{\tt garry@kemp.za.org} and
Hendrik J.R. Van Zyl$^{d,e}$\footnote{\tt hjrvanzyl@gmail.com}}}
\vspace{.8cm}
\centerline{{\it $^{a}$School of Science, Huzhou University, Huzhou 313000, China,}}
\vspace{.8cm}
\centerline{{\it $^{b}$School of Physics and Mandelstam Institute for Theoretical Physics,}}
\centerline{{\it University of the Witwatersrand, Wits, 2050, }}
\centerline{{\it South Africa,}}
\vspace{.8cm}
\centerline{{\it $^{c}$Department of Physics,}}
\centerline{{\it University of Johannesburg, Auckland Park, 2006, South Africa,}}
\vspace{.8cm}
\centerline{{\it $^{d}$The Laboratory for Quantum Gravity \& Strings,}}
\centerline{{\it Department of Mathematics \& Applied Mathematics,}}
\centerline{{\it University of Cape Town, Cape Town, South Africa,}}
\vspace{.8cm}
\centerline{{\it $^{e}$ The National Institute for Theoretical and Computational Sciences,}} \centerline{{\it Private Bag X1, Matieland, South Africa.}}

\vspace{1truecm}

\thispagestyle{empty}

\centerline{\bf ABSTRACT}

\vskip.2cm 

Bilocal holography provides a constructive approach to the vector model/higher spin gravity duality. It has two ingredients: a change of field variables and a change of space time coordinates. The change of field variables ensures that the loop expansion parameter becomes ${1\over N}$. The change of coordinates solves the Clebsch-Gordan problem of moving from the tensor product basis (in which the collective bilocal field is written) to the direct sum basis (appropriate for the description of the gravity fields). We argue that the change of space time coordinates can be deduced by requiring that operators constructed in the bilocal collective field theory are dual to local operators in the AdS bulk.

\setcounter{page}{0}
\setcounter{tocdepth}{2}
\newpage
\tableofcontents
\setcounter{footnote}{0}
\linespread{1.1}
\parskip 4pt

{}~
{}~

\section{Introduction}

Bilocal holography \cite{Das:2003vw} is a constructive approach to the duality \cite{Klebanov:2002ja,Sezgin:2002rt} between $O(N)$ vector models and higher spin gravity\cite{Vasiliev:1990en,Vasiliev:2003ev}. It is a concrete example of the general framework of collective field theory \cite{Jevicki:1979mb,Jevicki:1980zg}, which provides a constructive approach to the AdS/CFT duality \cite{Maldacena:1997re,Gubser:1998bc,Witten:1998qj} and other gauge theory/gravity dualities \cite{Polyakov:1998ju}. Concretely, bilocal holography gives an explicit formula for the higher spin gravity fields in terms of the operators of the conformal field theory. This realizes a construction of `precursors' i.e. bulk fields in terms of boundary operators \cite{Polchinski:1999yd}, as collective fields. The mapping between the boundary and bulk theories is achieved by matching the independent degrees of freedom in the conformal field theory to the independent degrees of freedom in the completely gauge fixed higher spin gravity. The fact that this can actually be carried out in complete detail is thanks to impressive work of Metsaev \cite{Metsaev:1999ui,Metsaev:2008fs,Metsaev:2008ks,Metsaev:2009hp,Metsaev:2011uy,Metsaev:2013wza} which achieves a complete gauge fixing (to light cone gauge) of the higher spin gravity and the reduction to independent degrees of freedom on both sides of the duality. 

Bilocal holography has two ingredients: there is a change of field variables and a change of space time coordinates. The change of field variables ensures that while the loop expansion parameter before transformation is $\hbar$, it becomes ${1\over N}$ after the
 transformation. The change of coordinates solves the Clebsch-Gordan problem of moving from the tensor product basis (in which the collective bilocal field is written) to the direct sum basis (appropriate for the gravity fields). This change of coordinates is highly non-trivial and the correctness of this step was verified by showing that the complete set of generators of the conformal field theory are mapped into those of the higher spin gravity \cite{deMelloKoch:2010wdf}. Bilocal holopgraphy has been developed in a number of interesting directions \cite{Jevicki:2011ss,Jevicki:2011aa,deMelloKoch:2012vc,deMelloKoch:2014mos,deMelloKoch:2014vnt,Mulokwe:2018czu,deMelloKoch:2018ivk,deMelloKoch:2021cni,deMelloKoch:2022sul,Johnson:2022cbe,deMelloKoch:2023ngh,deMelloKoch:2023ylr,Mintun:2014gua,Aharony:2020omh}.

Focus for now on the duality between CFT$_3$ and higher spin gravity on AdS$_4$. Before gauge fixing and reducing to independent degrees of freedom on both sides, the higher spin gravity includes a bulk scalar as well as a gauge field of every even integer spin, while the single trace primaries of the conformal field theory are a scalar primary of dimension $\Delta =1$ and a family of conserved spinning currents, of dimension $s+1$ and spin $s$ for every even integer $s$. After gauge fixing and reduction the complete set of conformal field theory degrees of freedom are packaged in a single O($N$) invariant equal time bilocal field, while on the gravity side one remains with two polarizations ($A^{XX\cdots X}$ and $A^{X\cdots XZ}$) of the spinning gauge with spin $2s$ for $s>0$, as well as the bulk scalar. Given that we consider a light cone gauge fixing, it is natural to pursue a light front quantization of both theories. In terms of the original scalar field $\phi^a$ of the O($N$) model, the bilocal field is written as
\bea
\sigma(x^+,x_1^-,x_1,x_2^-,x_2)&=&\phi^a(x^+,x^-_1,x_1)\phi^a(x^+,x^-_2,x_2)
\eea
where $x^\pm$ are the light cone coordinates of CFT$_3$ and $x$ is the single coordinate transverse to the light cone. The bilocal field develops a large $N$ expectation value, which we denote as $\sigma_0(x^+,x^-,x_1, x^-_2,x_2)$. Expanding about this background defines the fluctuation $\eta (x^+,x^-_1,x_1,x^-_2, x_2)$ as follows
\bea
\sigma(x^+,x^-_1,x_1,x^-_2, x_2)&=&\sigma_0(x^+,x^-_1,x_1,x^-_2, x_2)+\eta(x^+,x^-_1,x_1,x^-_2, x_2)
\eea
It is the fluctuation $\eta(x^+,x^-_1,x_1,x^-_2, x_2)$ that is identified with the dynamical fields of the dual higher spin gravity. This is a single field that depends on 5 coordinates. On the higher spin gravity side, it is useful to package the complete collection of higher spin gauge fields and the bulk scalar into a single field, with the help of a book keeping coordinate $\theta$, as follows
\bea
\Phi(X^+,X^-,X,Z,\theta)&=&\sum_{s=0}^\infty\Bigg({A^{XX\cdots XX}(X^+,X^-,X,Z)\over Z}\cos(2s\theta)\cr\cr
&&\qquad\quad +\,{A^{XX\cdots XZ}(X^+,X^-,X,Z)\over Z}\sin(2s\theta)\Bigg)
\eea
The fields on the right hand side are all tangent space fields. See \cite{Metsaev:1999ui} for a useful discussion. Following \cite{deMelloKoch:2021cni}, the mapping of the fields is most easily written in a mixed position/momentum space representation. In both the conformal field theory and in the higher spin gravity there is a symmetry under translations of $x^-$ and $X^-$ respectively. This allows us to perform a Fourier transform which trades $x_1^-$ and $x_2^-$ for $p_1^+$ and $p_2^+$ in the conformal field theory, and $X^-$ for $P^+$ in the higher spin gravity. The relation between the field can then be written as
\bea
\Phi(X^+,P^+,X,Z,\theta)&=&\mu(p_1^+,p_2^+)\eta(x^+,p_1^+,x_1,p_2^+,x_2)
\eea
where
\bea
\mu(p_1^+,p_2^+)&=&\sqrt{p_1^+p_2^+}
\eea
The relation between the coordinates of the conformal field theory and the bulk AdS$_4$ space time is given by $x^+=X^+$ as well as
\bea
P^+&=&p_1^++p_2^+\qquad X\,\,=\,\,{p_1^+ x_1+p_2^+x_2\over p_1^++p_2^+}\cr\cr
Z&=&{\sqrt{p_1^+p_2^+}\over p_1^++p_2^+}|x_1-x_2|\qquad\qquad
\theta\,\,=\,\,2\arctan \sqrt{p_2^+\over p_1^+}\label{coordtrans}
\eea

There are many tests that can be carried out once the bilocal holography mapping, as stated above, is given. For example, bilocal holography provides an explicit bulk reconstruction for complete set bulk gauge fields as well as the scalar field. This entails proving that the field $\Phi(X^+,P^+,X,Z,\theta)$ obeys the correct equation of motion \cite{deMelloKoch:2010wdf} as well as the correct GKPW boundary conditions as $Z\to 0$ \cite{Mintun:2014gua}. It is also interesting to explore whether or not information localizes in the bilocal holography construction as is expected in a theory quantum gravity. By restricting to a single subregion of the conformal field theory, one finds \cite{deMelloKoch:2021cni} that the subregion of the bulk that can be reconstructed is in perfect accord with entanglement wedge reconstruction \cite{Czech:2012bh,Headrick:2014cta,Wall:2012uf,Jafferis:2015del,Dong:2016eik,Cotler:2017erl}. Further by using the mapping it is possible to translate the statement of the holography of information \cite{Laddha:2020kvp,Chowdhury:2020hse,Raju:2020smc,Raju:2021lwh,SuvratYouTube} in the gravity theory into a statement in the conformal field theory. Using the usual operator product expansion formula the holography of information can be demonstrated directly in the conformal field theory \cite{deMelloKoch:2022sul}. 

Our goal in this paper is explore another feature of the holographic duality, namely bulk locality. We will see that constructing fields that are local in the bulk AdS spacetime naturally leads us to the coordinate map (\ref{coordtrans}) appearing in bilocal holography. This result which is the central result of this paper, is a useful observation since it outlines a deductive approach to determine the change of coordinates appearing in the bilocal holography map.

In what follows we consider the general case of the duality between CFT$_d$ and AdS$_{d+1}$\footnote{We limit ourselves to $d\ge 3$. The $d=2$ case is more subtle because in this case the free scalar fields are not primary operators.}. In this case the bilocal $\eta(x^+,x_1^- , \vec{x}_1,x_2^-,\vec{x}_2)$ is a function of $2d-1$ coordinates and the vector $\vec{x}$ is a $d-2$ dimensional vector transverse to the light cone. The bulk higher spin fields are again collected into a single field\footnote{We are using the very useful oscillator representation introduced in \cite{Metsaev:1999ui}.}
\bea
\Phi(X^+,X^-,\vec{X},Z,\alpha^i)&=&\sum_{s=0}^\infty\alpha_{a_1}\alpha_{a_2}\cdots \alpha_{a_{2s}}{A^{a_1 a_2\cdots a_{2s}}(X^+,X^-,\vec{X},Z)\over Z^{d-1\over 2}}|0\rangle
\eea
The index $a_i$ on the oscillators runs over $Z$ and the $d-2$ directions $\vec{X}$ transverse to the lightcone. Only $d-2$ of these oscillators are independent since we must impose that $\Phi(X^+,X^-,\vec{X},Z,\alpha^i)$ is traceless \cite{Mintun:2014gua}. There are $d+1$ coordinates $X^\pm,\vec{X},Z$ needed to specify an event in the AdS bulk and $d-2$ independent oscillators so that $\Phi(X^+,X^-,\vec{X},Z,\alpha^i)$ is a function of $2d-1$ coordinates. The holographic mapping between the fields in this case is
\bea
\Phi(X^+,P^+,\vec{X},Z,\alpha)&=&\mu(p_1^+,p_2^+,Z)\eta(x^+,p_1^+,\vec{x}_1,p_2^+,\vec{x}_2)
\eea
where
\bea
\mu(p_1^+,p_2^+,Z)&=&(p_1^+p_2^+)^{4-d\over 2}Z^{3-d\over 2}(P^+ Z)^{d-3}
\eea
In the rest of the article we will determine the relation between the coordinates of the $d$ dimensional bilocal field theory ($x^+,p_1^+,\vec{x}_1,p_2^+,\vec{x}_2$) and those of the bulk AdS$_{d+1}$ space time ($X^+,P^+,\vec{X},Z,\alpha^i$), as well as the form for $\mu(p_1^+,p_2^+,Z)$ quoted above.  This coordinate mapping will be determined by requiring that the bilocal field of the conformal field theory is dual to a local operator in the bulk.

\section{Bulk Locality}\label{bulklocality}

A point in the bulk AdS$_{d+1}$ space time is specified by $d+1$ coordinates\footnote{In what follows we always use capital letters for coordinates of the bulk AdS spacetime and little letters for the coordinates of the space time of the CFT.}  $X^M=(X^\mu,Z)$, where $X^\mu$ is a set of $d$ coordinates and $Z$ is the radial (holographic) coordinate. The bulk metric is
\bea
   ds^2&=&{dX^\mu dX^\nu\eta_{\mu\nu}+dZ^2\over Z^2}
\eea
where $\eta={\rm diag}(-1,1,\cdots,1)$. The isometries of this metric are generated by
\bea
P^\mu &=&\partial^\mu\qquad\qquad\qquad\qquad
J^{\mu\nu}\,\,=\,\,X^\mu\partial^\nu-X^\nu\partial^\mu\cr\cr
D&=&X^\mu\partial_\mu+Z\partial_Z\qquad\qquad
K^\mu\,\,=\,\,-{1\over 2}(X^\mu X_\mu+Z^2)\partial^\mu+X^\mu D
\eea
These generators generate the group SO(2,$d$). In what follows we employ lightcone coordinates, obtained by setting
\bea
X^\pm &=& {1\over\sqrt{2}}\left(X^{d-1}\pm X^0\right)
\eea
The vector $\vec{X}$ has components $X^i$ for $i=1,2,\cdots,d-2$ which are coordinates transverse to the lightcone. $Z$ is the coordinate for the extra holographic dimension.

AdS spacetime is an example of a symmetric space\cite{symspace}. Starting from any particular point in the AdS spacetime, we can reach any other point with the action of some element of SO(2,$d$). Not every element of SO(2,$d$) will move a particular point: each point has an isotropy group under whose action the given point is fixed. For the case we are discussing here, the isotropy group is SO(1,$d$) so that we can identify AdS spacetime as the coset SO(2,$d$)/SO(1,$d$). As a particularly simple example, consider the bulk point $X^\mu=0$ and $Z=Z_p$ where $Z_p$ is a definite value. Fixing these values of the coordinates the generators become
\bea
P^\mu&=&\partial^\mu\qquad J^{\mu\nu}\,\,=\,\,0\qquad D=Z_p\partial_Z\qquad K^\mu\,\,=\,\,-{1\over 2}Z_p^2\partial^\mu
\eea
Thus, the generators $J^{\mu\nu}$ and $K^\mu+{1\over 2}Z_p^2 P^\mu$ both vanish when acting on our bulk point. They are non-zero acting on any other point so we can define this bulk location as the point that is annihilated by the group generated by $\{J^{\mu\nu},K^\mu+{1\over 2}Z_p^2 P^\mu\}$. This group is SO(1,$d$) - it is the isotropy group of the bulk point we are considering. Note that the coset SO(2,$d$)/SO(1,$d$) is a $d+1$ dimensional space and each class in this coset corresponds to a point in the bulk of the AdS$_{d+1}$ spacetime. Choosing a different bulk point leads to a different isotropy group, which is still isomorphic to SO(1,$d$). The data of the isotropy group and the bulk point are equivalent pieces of data. The isotropy group is giving us information about bulk locality - it tells us what it means to be localized at a specific bulk point.

Since the isotropy group leaves the bulk point fixed, it will only shuffle the different polarizations of a spinning field amongst each other. Thus, for example, to construct an operator $O_\Psi$ in the bilocal field theory that is dual to a bulk operator localized at $X^M=(0,Z_p)$ in the AdS bulk, we solve the equations\footnote{Here we are studying a scalar state for simplicity. If we have a spinning state we would again use oscillators and add the spin contribution $M^{\mu\nu}=\alpha^\mu\bar{\alpha}^\nu-\alpha^\nu\bar{\alpha}^\mu$ to $J^{\mu\nu}$ and $M^{\mu\nu}x_\nu$ to $K^\mu$. In addition $J^{\mu\nu}$ would no longer annihilate the state, but would mix the different spin states amongst each other.}
\bea
\big[\,J^{\mu\nu}\,,\,O_\Psi\big]&=&0\,\,=\,\,\big[ K^\mu+{1\over 2}Z_p^2 P^\mu\,,\,O_\Psi\big]\label{blkcndtn}
\eea
This is the minimal requirement for operators in the bilocal field theory to be dual to operators localized in the AdS bulk. See \cite{Miyaji:2015fia,Verlinde:2015qfa,Nakayama:2015mva,Kabat:2017mun} for related discussion. The second equation above is telling: an operator localized at a boundary point is primary, whilst an operator located in the bulk corresponds to a non-trivial combination of the primary and its descendents. So AdS/CFT geometrizes the space of CFT operators, introducing an extra holographic dimension with coordinate $Z$ and placing the primaries on the boundary of the AdS bulk. Since the single trace primaries generate the complete set of gauge invariant operators, all the information sits on the boundary, consistent with the holography of information.

Our goal is now to use the requirement of bulk locality to derive the coordinate transformation of bilocal holography. Towards that end, consider the set of bulk points given by $X^+=0$, $\vec{X}=\vec{X}_p$, $Z=Z_p$ where $\vec{X}_p$ and $Z_p$ are definite values. Notice that $X^-$ is left arbitrary i.e. we are considering a light like line of points in the AdS$_{d+1}$ bulk. For this line of points the momentum $P^+$ and special conformal transformation $K^+$ become 
\bea
P^+&=&\partial^+\qquad\qquad K^+\,\,=\,\,-{1\over 2}(\vec{X}\cdot\vec{X}+Z_p^2)\partial^+
\eea
The second of (\ref{blkcndtn}) provides a non trivial differential equation that a bulk field localized on the light like line must obey
\bea
\big( K^++{1\over 2}(\vec{X}\cdot\vec{X}+Z_p^2)P^+\big)O_\Psi=0\label{exprsblkloc}
\eea
This equation is completely general, holding for fields of any spin, thanks to the form of $K^+$ given in (3.71) of \cite{Metsaev:1999ui}. The coordinate transformation (\ref{coordtrans}) was determined by matching the generators of conformal transformations of the bilocal conformal field theory with those of the higher spin gravity. To obtain a relation between the coordinates of the field theory and those of the gravity, we will insert the bilocal generators into the above equation. Further, if we now take a Fourier transform on $X^-$ in gravity, the differential operator $\partial^+$ is replaced by the variable $P^+$. Since we have localized only to the light like line parameterized by $X^-$, the Fourier transformed field obeys the same bulk locality condition. Similarly, if we take a Fourier transform in the bilocal theory, $\partial^+_1$ and $\partial^+_2$ are replaced by $p^+_1$ and $p^+_2$. Bilocal holography matches the Hilbert space of the bilocal theory with that of the higher spin gravity. Since these are each defined at fixed light cone time we should identify $X^+=x^+$. Inserting the bilocal expression for the generators into (\ref{exprsblkloc}), we obtain
\bea 
\Big(-{1\over 2}(\vec{x}_1\cdot\vec{x}_1 p^+_1+\vec{x}_2\cdot\vec{x}_2 p^+_2)+{1\over 2}(\vec{X}\cdot\vec{X}+Z_p^2)(p_1^++p_2^+)\Big)O_\Psi&=&0
\eea
Notice that this is a polynomial multiplied by the field. Since the field does not vanish the polynomial does and in the end we obtain the equation
\bea
-{1\over 2}(\vec{x}_1\cdot\vec{x}_1 p^+_1+\vec{x}_2\cdot\vec{x}_2 p^+_2)+{1\over 2}(\vec{X}_p\cdot\vec{X}_p+Z_p^2)(p_1^++p_2^+)&=&0
\label{eqntosolve}
\eea
which relates the bulk AdS$_{d+1}$ coordinates $\vec{X}_p$ and $Z_p$ to the coordinates
$\vec{x}_1$, $\vec{x}_2$, $p_1^+$ and $p_2^+$ of the conformal field theory.  

\section{Holographic Mapping}\label{hmap}

In this section we will solve (\ref{eqntosolve}) to explicitely demonstrate the link between bulk locality and the bilocal holography mapping. One of the key pieces of evidence motivating the discovery of AdS/CFT was a matching between the global symmetries of the conformal field theory and the isometries of the dual AdS gravity. Our solution of (\ref{eqntosolve}) will also make use of this fact.

The symmetry $X^-\to X^-+a$ in the bulk corresponds to $x_1^-\to x_1^-+a$ and $x_2^-\to x_2^-+a$ in the bilocal collective field theory. Thus, the generator producing an infinitesimal transformation of $X^-$ in the bulk must generate an infinitesimal transformation of both $x_1^-$ and $x_2^-$ which implies that
\bea
P^+&=&p_1^++p_2^+
\eea
An identical argument for the directions transverse to the light cone allows us to conclude that
\bea
P^i=p_1^i+p_2^i
\eea
Now consider an SO($d$-2) transformation $R^i{}_j$ which acts as
\bea
X^i\to R^i{}_j X^j
\eea
in the bulk, and as 
\bea
x_1^i\to R^i{}_j x_1^j\qquad x_2^i\to R^i{}_j x_2^j
\eea
in the bilocal field theory. Thus, $\vec{X}$, $\vec{x}_1$ and $\vec{x}_2$ are all in the $d-2$ dimensional vector representation of SO($d$-2). Higher powers of these coordinates are not generally even in an ireducible representation, and in particular, they are not in the vector representation. A simple way to ensure that all three are in the vector representation is to take a linear relation between them\footnote{In principle one could add a constant $\vec{X}_0$ on the right hand side. This can always be removed with a judicious choice of origin.}
\bea
\vec{X}=\alpha\vec{x}_1+\beta\vec{x}_2
\eea
Next, the translation $\vec{X}\to\vec{X}+\vec{a}$ in the bulk corresponds to $\vec{x}_1\to\vec{x}_1+\vec{a}$ and $\vec{x}_2\to\vec{x}_2+\vec{a}$ in the bilocal field theory. This forces $\beta+\alpha=1$ so that we have
\bea
\vec{X}=\alpha\vec{x}_1+(1-\alpha)\vec{x}_2\label{formofX}
\eea
The arguments we have considered so far allow the parameter $\alpha$ to be an arbitrary function of $p_1^+$ and $p_2^+$.

Under a translation $X^i\to X^i+a^i$ the bulk coordinate $Z$ is unchanged. This translation takes $x_1^i\to x_1^i+a^i$ and $x_2^i\to x_2^i+a^i$ in the bilocal field theory so that $Z$ is a function only of $x_1^i-x_2^i$. Next, under the SO($d$-2) transformation $R^i{}_j$ we know that $Z$ is unchanged. This implies that $Z$ is a function only of $|\vec{x}_1-\vec{x}_2|$. If we now assume that $Z$ is a linear function of $|\vec{x}_1-\vec{x}_2|$ we can write
\bea
Z&=& \delta |\vec{x}_1-\vec{x}_2|
\eea
where again, our arguments allow $\delta$ to be an arbitrary function of $p_1^+$ and $p_2^+$. The assumption of linearity will be motivated below.

Recall that the equation we wish to solve, (\ref{eqntosolve}), is given by
\bea
-(\vec{x}_1\cdot\vec{x}_1p_1^++\vec{x}_2\cdot\vec{x}_2p_2^+)+(\vec{X}\cdot\vec{X}+Z^2)(p_1^++p_2^+)&=&0\label{inhere}
\eea
A simple computation shows that
\bea
\vec{X}\cdot\vec{X}+Z^2&=& \alpha^2\vec{x}_1\cdot\vec{x}_1+(1-2\alpha+\alpha^2)\vec{x}_2\cdot\vec{x}_2+2\alpha(1-\alpha)\vec{x}_1\cdot\vec{x}_2\cr\cr
&&\,\,+\,\,\delta^2(\vec{x}_1\cdot\vec{x}_1+\vec{x}_2\cdot\vec{x}_2-2\vec{x}_1\cdot\vec{x}_2)
\label{usethis}
\eea
Inserting (\ref{usethis}) into (\ref{inhere}) and equating the coefficients of $\vec{x}_1\cdot\vec{x}_1$, $\vec{x}_2\cdot\vec{x}_2$ and $\vec{x}_1\cdot\vec{x}_2$ to zero, we obtain the following three equations
\bea
- p_1^++(\alpha^2+\delta^2)(p_1^++p_2^+)&=&0\cr\cr
- p_2^++(1-2\alpha+\alpha^2+\delta^2)(p_1^++p_2^+)&=&0\cr\cr
(2\alpha(1-\alpha)-2\delta^2)(p_1^++p_2^+)&=&0
\eea
The last equation above implies that $\delta^2=\alpha(1-\alpha)$. Inserting this into the first equation above gives a linear equation for $\alpha$ so that $\alpha$ and $\delta$ are determined. This unique solution ensures that all three equations above are satisfied. The solution is
\bea
\alpha&=&{p_1^+\over p_1^++p_2^+}\qquad\qquad
1-\alpha\,\,=\,\,{p_2^+\over p_1^++p_2^+}\qquad\qquad
\delta\,\,=\,\,{\sqrt{p_1^+p_2^+}\over p_1^++p_2^+}
\eea
This argument therefore implies that
\bea
\vec{X}&=&{p_1^+\vec{x}_1+p_2^+\vec{x}_2\over p_1^++p_2^+}
\qquad
Z\,\,=\,\,{\sqrt{p_1^+p_2^+}\over p_1^++p_2^+}|\vec{x}_1-\vec{x}_2|
\eea
which is precisely the coordinate transformation for the bulk AdS coordinates given by the bilocal holography mapping (\ref{coordtrans}).

Let us now return to the assumption that $Z$ is linear in $|\vec{x}_1-\vec{x}_2|$. To get some insight into why this must be the case, note that we can rewrite (\ref{eqntosolve}) as
\bea
-(\vec{x}_1\cdot\vec{x}_1p_1^++\vec{x}_2\cdot\vec{x}_2p_2^+)+\vec{X}\cdot\vec{X}(p_1^++p_2^+)&=&-Z^2(p_1^++p_2^+)
\eea
Recalling the formula (\ref{formofX}), it is clear that all terms on the left hand side of this equation are quadratic in $\vec{x}_i$. Thus, the right hand side must be too and this forces $Z$ to be a linear function of $|\vec{x}_1-\vec{x}_2|$.

An alternative way to approach (\ref{eqntosolve}) is to take its commutator with $P^i$, which gives
\bea
0 &=& -{1\over 2}\left([P^i,\vec{x}_1\cdot \vec{x}_1]p_1^++[P^i,\vec{x}_2\cdot\vec{x}_2] p_2^+\right)+{1\over 2}\left([P^i,\vec{X}_p\cdot\vec{X}_p]+[P^i, Z_p^2]\right)(p_1^+ + p_2^+)\cr\cr
&=&-x_1^i p_1^+ - x_2^i p_2^++X_p^i (p_1^+ + p_2^+)
\eea
which immediately implies that
\bea
X_p^i & = & \frac{x_1^i p_1^{+} + x_2^{i} p_2^{+}}{p_1^{+} + p_2^{+}} 
\eea
Inserting this into (\ref{eqntosolve}) we find
\bea
Z_p^2&=&\frac{(\vec{x}_1\cdot\vec{x}_1 p_1^++ \vec{x}_2\cdot\vec{x}_2 p_2^+)}{p_1^++p_2^+}-\vec{X}_p\cdot\vec{X}_p\cr\cr
&=&\frac{p_1^+ p_2^+}{(p_1^+ + p_2^+)^2} |\vec{x}_1 - \vec{x}_2|^2
\eea
which is the result we obtained above.

To complete the holographic mapping we still have to determine the angles that are used to package the higher spin fields into a single field. Following \cite{Jevicki:2011ss} we will show how the spin components of the angular momentum generators are determined. We start by considering the rotations transverse to the light cone\footnote{This formula assumes that $d>3$.}
\bea
J^{ij}&=&X^i P^j-P^i X^j+m^{ij}
\eea
The first two terms are the orbital contribution while the third term in the spin part. It is the spin part that we would like to determine. It is therefore convenient to perform the analysis at the bulk point $X^i=0$, which sets the orbital terms to zero. In this case we have
\bea
J^{ij}&=&m^{ij}\label{spinpart}
\eea
This should be matched to the generator of the bilocal field theory, which reads
\bea
J^{ij}=x_1^i{\partial\over\partial x_1^j}-x_1^j{\partial\over\partial x_1^i}+x_2^i{\partial\over\partial x_2^j}-x_2^j{\partial\over\partial x_2^i}\label{cftjij}
\eea
To perform the comparison, note that since the angles are translation invariant, they can only be a function only of $x_1^i-x_2^i$. Setting $X^i=0$ implies that $\vec{x}_1$ and $\vec{x}_2$ are not independent variables, but rather they obey the relation
\bea
p_1^+\vec{x}_1+p_2^+\vec{x}_2=0
\eea
Using this relation, it is simple to argue that
\bea
\vec{x}_1={p_2^+\over p_1^++p_2^+}(\vec{x}_1-\vec{x}_2)\qquad\qquad
\vec{x}_2=-{p_1^+\over p_1^++p_2^+}(\vec{x}_1-\vec{x}_2)
\eea
After using these identities in (\ref{cftjij})  and equating the result to (\ref{spinpart}) we learn that
\bea
m^{ij}&=&{(x_1-x_2)^i\over p_1^++p_2^+}(p_2^+p_1^j-p_1^+p_2^j)-{(x_1-x_2)^j\over p_1^++p_2^+}(p_2^+p_1^i-p_1^+p_2^i)
\eea
which is in complete agreement with the known result \cite{Jevicki:2011ss}.

Next consider the $m^{iz}$ contribution to the spin angular momentum. It proves useful to study the special conformal generator $K^i$ which is given by \cite{Metsaev:1999ui}
\bea
K^i=-{1\over 2}\left(2X^+X^-+\vec{X}\cdot\vec{X}+Z^2\right)P^i+X^i D+m^{ij}X^j+m^{iZ}Z+m^{i-}X^+
\eea
This can be simplified dramatically if we situate ourselves at the light like line of bulk points specified by $X^+=0=X^i$ and any $X^-$. In this case $K^i$ becomes
\bea
K^i&=&-{1\over 2}Z^2 P^i+m^{iZ}Z\label{BlkKi}
\eea
The only unknown in this expression is $m^{iZ}$. We know that the spin generator $m^{iZ}$ is translation invariant which implies that it is again only a function only of $x_1^i-x_2^i$. Equating this to the corresponding expression in the bilocal field theory, we obtain
\bea
K^i&=& \mu(p_1^+,p_2^+,Z)\left(-{1\over 2}(\vec{x}_1\cdot\vec{x}_1p_1^i+\vec{x}_2\cdot\vec{x}_2p_2^i)+x_1^i D_1+ x_2^i D_2\right)\frac{1}{\mu(p_1^+,p_2^+,Z)}\cr\cr
&=&\mu(p_1^+,p_2^+,Z)\Big(-{1\over 2}\left({(p_2^+)^2\over (p_1^++p_2^+)^2}|\vec{x}_1-\vec{x}_2|^2 p_1^i+{(p_1^+)^2\over (p_1^++p_2^+)^2}|\vec{x}_1-\vec{x}_2|^2 p_2^i\right)\cr\cr
&&\qquad +{p_2^+\over p_1^++p_2^+}(x_1^i-x_2^i) D_1-{p_1^+\over p_1^++p_2^+}(x_1^i-x_2^i) D_2\Big)\frac{1}{\mu(p_1^+,p_2^+,Z)}\label{bndryKi}
\eea
where when acting on $\phi^a(x^+,x^-_1,x_1)$ we have
\bea
D_1&=&x^+ p_1^-+x_1^- p_1^++x_1^j p_1^j+{d-2\over 2}\cr\cr
&=& x_1^- p_1^++{p_2^+\over p_1^++p_2^+}(x_1^j-x_2^j) p_1^j+{d-2\over 2}
\eea
and when acting on $\phi^a(x^+,x^-_2,x_2)$ we have
\bea
D_2 &=& x_2^- p_2^+-{p_1^+\over p_1^++p_2^+}(x_1^j-x_2^j) p_2^j+{d-2\over 2}
\eea
Equating (\ref{BlkKi}) and (\ref{bndryKi}) we easily find
\bea
m^{iZ}&=&{x_1^i-x_2^i\over |\vec{x}_1-\vec{x}_2|}\left[\sqrt{p_1^+p_2^+}(x_1^--x_2^-)+{((p_1^+)^2 p_2^j+(p_2^+)^2p_1^j)(x_1^j-x_2^j)\over (p_1^++p_2^+)\sqrt{p_1^+p_2^+}}\right]\cr\cr
&&+\,{1\over 2}\,\,{p_1^+-p_2^+\over p_1^++p_2^+}\,|\vec{x}_1-\vec{x}_2|\,\left(p_1^i\sqrt{p_2^+\over p_1^+}-p_2^i\sqrt{p_1^+\over p_2^+}\right)
\eea
which is in complete agreement with the known result \cite{Jevicki:2011ss,Mintun:2014gua}. This computation also fixes
\bea
\mu(p_1^+,p_2^+,Z)=f(Z,P^+)(p_1^+p_2^+)^{4-d\over 2}
\eea
Finally, by matching the action of the CFT dilatation operator on $\mu(p_1^+,p_2^+,Z)\eta(x^+,p_1^+,\vec{x}_1,p_2^+,\vec{x}_2)$ with the action of the bulk dilatation operator on $\Phi(X^+,P^+,\vec{X},Z,\alpha)$, we fix
\bea
\mu(p_1^+,p_2^+,Z)=(p_1^+p_2^+)^{4-d\over 2}Z^{3-d\over 2}(P^+ Z)^{d-3}
\eea

This completes the derivation of the coordinate transformation, by making use of bulk locality.

\section{A concrete example}\label{example}

In the previous section we have written down the generators of the angular momentum. In this section, for the specific case that $d=4$, we will translate these angular momenta into a collection of angles. Denote the coordinates of the conformal field theory by $x^\mu=(t,w,x,y)$. Light cone coordinates are defined as $x^\pm = t\pm w$ and the bilocal field depends on 7 coordinates $\sigma(x^+,x_1^-,x_1,y_1,x_2^-,x_2,y_2)$. Bilocal holography relates these 7 coordinates to the AdS$_5$ coordinates $X^+,X^-,X,Y,Z$ and two angles $\theta,\varphi$. The two angles will be extracted from the angular momenta $m^{XY}$, $m^{XZ}$ and $m^{YZ}$.

Using the results of the previous section, we have
\bea
m^{XY}&=&(x_1-x_2){p_2^+p_1^y-p_1^+p_2^y\over p_1^++p_2^+} -(y_1-y_2){p_2^+p_1^x-p_1^+p_2^x\over p_1^++p_2^+}
\eea
Now, to interpret this formula note that
\bea
\left[{p_2^+p_1^i-p_1^+p_2^i\over p_1^++p_2^+},{p_1^+ x_1^j+p_2^+x_2^j\over p_1^++p_2^+}\right]&=&0\qquad\qquad
\left[{p_2^+p_1^i-p_1^+p_2^i\over p_1^++p_2^+},x_1^j-x_2^j\right]\,\,=\,\,\delta^{ij}
\eea
so that we can interpret the momentum ${\cal P}^i={p_2^+p_1^i-p_1^+p_2^i\over p_1^++p_2^+}$ as the momentum conjugate to the relative coordinate. The total momentum, which must commute with the relative coordinate and is conjugate to the $\vec{X}$ coordinate, also behaves as expected
\bea
\left[p_1^i+p_2^i,x_1^j-x_2^j\right]&=&0\qquad
\left[p_1^i+p_2^i,{p_1^+ x_1^j+p_2^+x_2^j\over p_1^++p_2^+}\right]\,\,=\,\,\delta^{ij}
\eea
Consequently, if we set
\bea
x_1-x_2&=& |\vec{x}_1-\vec{x}_2| \cos (\varphi)\qquad\qquad
y_1-y_2\,\,=\,\, |\vec{x}_1-\vec{x}_2| \sin (\varphi)
\eea
where
\bea
|\vec{x}_1-\vec{x}_2|\equiv \sqrt{(x_1-x_2)^2-(y_1-y_2)^2}
\eea
then we have
\bea
m^{XY}&=&{\partial\over\partial\varphi}
\eea
Next consider the pair of generators
\bea
m^{XZ}&=&{x_1-x_2\over |\vec{x}_1-\vec{x}_2|}\left[\sqrt{p_1^+p_2^+}(x_1^--x_2^-)+{((p_1^+)^2 p_2^x+(p_2^+)^2p_1^x)(x_1-x_2)+((p_1^+)^2 p_2^y+(p_2^+)^2p_1^y)(y_1-y_2)\over (p_1^++p_2^+)\sqrt{p_1^+p_2^+}}\right]\cr\cr
&&+\,{1\over 2}\,\,{p_1^+-p_2^+\over p_1^++p_2^+}\,|\vec{x}_1-\vec{x}_2|\,\left(p_1^x\sqrt{p_2^+\over p_1^+}-p_2^x\sqrt{p_1^+\over p_2^+}\right)\cr\cr
&=&\cos\varphi{\partial\over\partial\theta}-\cot\theta\sin\varphi{\partial\over\partial\varphi}\cr\cr
&=&\cos\varphi P_\theta-\cot\theta\sin\varphi P_\varphi
\eea
\bea
m^{YZ}&=&{y_1-y_2\over |\vec{x}_1-\vec{x}_2|}\left[\sqrt{p_1^+p_2^+}(x_1^--x_2^-)+{((p_1^+)^2 p_2^x+(p_2^+)^2p_1^x)(x_1-x_2)+((p_1^+)^2 p_2^y+(p_2^+)^2p_1^y)(y_1-y_2)\over (p_1^++p_2^+)\sqrt{p_1^+p_2^+}}\right]\cr\cr
&&+\,{1\over 2}\,\,{p_1^+-p_2^+\over p_1^++p_2^+}\,|\vec{x}_1-\vec{x}_2|\,\left(p_1^y\sqrt{p_2^+\over p_1^+}-p_2^y\sqrt{p_1^+\over p_2^+}\right)\cr\cr
&=&\sin\varphi{\partial\over\partial\theta}+\cot\theta\cos\varphi{\partial\over\partial\varphi}\cr\cr
&=&\sin\varphi P_\theta+\cot\theta\cos\varphi P_\varphi
\eea
where the angles $\theta,\varphi$ are given by
\bea
\theta\,\,=\,\,2\arctan \sqrt{p_2^+\over p_1^+}\qquad\qquad
\varphi\,\,=\,\,\arctan {y_1-y_2\over x_1-x_2}
\eea
which agrees with the angle $\theta$ appearing in the $d=3$ bilocal holography map. Thus, we are indeed able to identify two extra angles. Finally, the momenta conjugate to these angles are
\bea
P_\varphi&=&m^{XY}\cr\cr
P_\theta&=&\sqrt{p_1^+p_2^+} (x_1^- -x_2^-)+{x_1 - x_2\over 2}\left( \sqrt{p_2^+\over p_1^+}p_1^x + \sqrt{p_1^+\over p_2^+}p_2^x\right)+{y_1-y_2\over 2}\left( \sqrt{p_2^+\over p_1^+}p_1^y+\sqrt{p_1^+\over p_2^+}p_2^y\right)\cr\cr
&&
\eea

\section{Discussion and Conclusions}\label{conclusions}

Collective field theory provides a constructive approach to the AdS/CFT duality and other gauge theory/gravity dualities. It formulates the dynamics of quantum field theory in terms of gauge invariant collective fields. This reorganization of the degrees of freedom has an important consequence: while the loop expansion parameter of the original theory is $\hbar$, after the change of field variables the loop expansion parameter becomes ${1\over N}$, matching that of the dual gravity description. For the specific case of the duality between $O(N)$ vector models and higher spin gravity, the collective fields are bilocal and the resulting collective construction is called bilocal holography.

Bilocal holography has two ingredients: the change of field variables we have just discussed, as well as a change of space time coordinates. The change of space time coordinates is needed to develop the physical interpretation of the theory. The scalar field of the O($N$) model transforms in the spin zero and dimension ${d-2\over 2}$ representation of SO(2,$d$). The bilocal transforms in a tensor product of two copies of this representation. This representation is reducible and each irreducible component corresponds to a different bulk field in the gravitational description. Consequently, to develop the physical interpretation of the bilocal collective field theory we must solve the Clebsch-Gordan problem of moving from the tensor product basis (in which the collective bilocal field is written) to the direct sum basis (appropriate for the gravity fields). This is accomplished by a change of space time coordinates. 

In the original work \cite{deMelloKoch:2010wdf} this change of coordinates was discovered by brute force and then verified by demonstrating that it reproduces the bulk AdS isometry generators starting from the conformal generators in the bilocal field theory. In this paper we have outlined a deductive approach to determining this change of coordinates. 

The idea is to formulate a minimal requirement for operators in the bilocal field theory to be dual to an operator located at a given bulk point. More correctly, we have considered the conditions needed to localize operators to a light like line in space time. The first step is the choice of the light like line in the bulk AdS space time. The isotropy group of this line is then determined. The generators of the isotropy group act by shuffling polarizations of the spinning field. Using a  Fourier transform on the light like line, we have constructed a mixed momentum/position description. Within this description an algebraic equation was determined, which fixed the form of the AdS bulk coordinates in terms of those of the bilocal conformal field theory. Finally, additional angles used to package the complete collection of spinning bulk fields into a single field, were determined (implicitly) by evaluating the spin contribution to the conformal generators.

There are two different approaches which can be followed to rewrite the vector model in terms of bi-local fields which are then mapped to the bulk. The approach taken in this article uses the Hamiltonian language. In this description the bi-local operators are constructed from fields at different points in space but at the same time. This breaks manifest Lorentz invariance. A second approach \cite{deMelloKoch:1996mj,deMelloKoch:2018ivk,Aharony:2020omh} uses bilocal fields constructed from fields that are separated both in space and time. This preserves Lorentz invariance as well as the full conformal group. Both descriptions are useful. The two time description is useful as it makes the underlying conformal symmetry manifest. On the other hand, studying the theory at finite temperature is straightforward in the Hamiltonian approach, but it is non-trivial in the two time approach. The relation between the single time and two time descriptions has been considered in \cite{Kamimura:1977dv,Jevicki:2011ss}. In the two-time description one can impose constraints and perform a gauge fixing to the single time description.

Our construction is likely to be useful in more general applications of collective field theory to gauge theory/gravity dualities. An immediate application would be to gauge theories. In this case, the theory includes matrix fields transforming in the adjoint representation. Consequently, there is a much richer set of invariant variables: one can take traces of fields at different locations, so that bilocal operators, trilocal operators and in general $k$-local operators appear in the set of invariants. The construction of the change of coordinates needed for this case is highly non-trivial and has been an obstacle to progress. The application of bulk locality can be used to overcome this difficulty \cite{us}. Another interesting application of bulk locality is to consider fermionic vector models \cite{Das:2012dt} which are potentially relevant for an understanding of the holography of de Sitter space. This application would involve using the isotropy group of a point in de Sitter spacetime.

\begin{center} 
{\bf Acknowledgements}
\end{center}
RdMK thanks Pawel Caputa for useful discussions on the subject of this paper. This research is supported by a start up research fund of Huzhou University, a Zhejiang Province talent award and by a Changjiang Scholar award. The author would like to thank the Isaac Newton Institute for Mathematical Sciences for support and hospitality during the programme ``Black holes: bridges between number theory and holographic quantum information'' when work on this paper was initiated. This work was supported by EPSRC Grant Number EP/R014604/1. HJRVZ is supported in part by the “Quantum Technologies for Sustainable Development” grant from the National Institute for Theoretical and Computational Sciences of South Africa (NITHECS).

\end{document}